\documentclass[letterpaper]{article}

\usepackage{xparse}
\NewDocumentCommand{\citename}{o m}{\citeauthor{#2}~\IfNoValueTF{#1}{\shortcite{#2}}{\shortcite[#1]{#2}}}

\usepackage{iccc}
\usepackage{times}
\usepackage{helvet}
\usepackage{courier}
\usepackage{color}
\usepackage{url}

\newcommand\EatDot[1]{}

\usepackage[framemethod=tikz]{mdframed}
\mdfsetup{
skipabove=\baselineskip,
skipbelow=\baselineskip,
innertopmargin=3pt,
innerbottommargin=8pt,
apptotikzsetting={\tikzset{mdfbackground/.append style={fill=white,fill opacity=0}}}
}

\usepackage[section]{placeins}

\usepackage{xifthen}

\newcommand{\Press}[1][]{%
\ifthenelse{\equal{#1}{}}{{\sf Place}}{{\sf Place#1}}%
}

\newcommand{\Place}[1][]{%
\ifthenelse{\equal{#1}{}}{{\sf Place}}{{\sf Place#1}}%
}

\newcommand{\Producer}[1][]{%
\ifthenelse{\equal{#1}{}}{{\sf Producer}}{{\sf Producer#1}}%
}

\newcommand{\Person}[1][]{%
\ifthenelse{\equal{#1}{}}{{\sf Producer}}{{\sf Producer#1}}%
}

\newcommand{\Process}[1][]{%
\ifthenelse{\equal{#1}{}}{{\sf Process}}{{\sf Process#1}}%
}

\newcommand{\Product}[1][]{%
\ifthenelse{\equal{#1}{}}{{\sf Product}}{{\sf Product#1}}%
}

\title{An institutional approach to computational social creativity}
\author{Joseph Corneli\\
Computational Creativity Group\\
Goldsmiths College\\
London, UK\\
j.corneli@gold.ac.uk\\
}
\usepackage{threeparttable}

\begin{document} 
\maketitle
\begin{abstract}
  Modelling the creativity that takes place in social settings
  presents a range of theoretical challenges.  Mel Rhodes's classic
  ``4Ps'' of creativity, the ``Person, Process, Product, and Press,''
  offer an initial typology.  Here, Rhodes's ideas are connected with
  Elinor Ostrom's work on the analysis of economic governance to
  generate several ``creativity design principles.''  These principles
  frame a survey of the shared concepts that structure the contexts
  that support creative work.  The concepts are connected to the idea
  of computational ``tests'' to foreground the relationship with
  standard computing practice, and to draw out specific
  recommendations for the further development of computational
  creativity culture.
\end{abstract}

\section{Introduction}

One two-part claim is advanced and defended herein: \emph{Elinor
  Ostrom's theory of institutions can be used to design systems that
  exhibit computational social creativity, and a culture supports this
  work}.  The contribution takes the form of several candidate
``design principles,'' a literature survey that elaborates them, and
an analysis that connects these ideas to common programming practice.

The paper is structured as follows.
The ``Background'' section describes Ostrom's
\shortcite{elinor1990governing} \emph{Institutional
Analysis and Development} (IAD) framework, focusing on her proposed design principles for commons
management.  To connect these ideas to social creativity, the
paper draws on the 4Ps (Person\slash Process\slash Press\slash
Product), a model for thinking about creative contexts
\cite{rhodes1961analysis} that has been brought to bear in
theorising computational creativity \cite{jordanous2016four}.  This is summarised and slightly adapted.
In the subsequent main section of the paper, ``Testing for
Creativity'', Ostrom's design principles are transposed from the world
of commons management to the world of computational social creativity.
This section looks for ways to connect the propsed creativity design
principles to computational methods, and also draws on contemporary
thinking in the philosophy of technology, with examples from familar
social computing settings like Wikipedia.
A two-part example dealing with both the ``soft'' culture of the
computational creativity community and potential software-based
interventions is presented in the ``Example'' section.
Finally, the ``Discussion and Conclusions'' highlight the relevance of
this work for computational creativity culture, systems, and evaluation.

\section{Background}

This section summarises the motivation for the paper, introduces
Elinor Ostrom's work, and reviews Rhodes's 4P framework.  
The central parts of this section
are Table \ref{tab:design-principles} and
\ref{tab:creativity-design-principles}, which list Ostrom's design
principles for managing a commons, and transpose them to creative
domains.

\paragraph{Motivation}

The current investigation is motivated, in part, by the idea of
\emph{Ecologically Grounded Creative Practice}
\cite{keller2014ubimus}.  Within a given ecological niche, agents and
objects interact; niches can also be brought into relationship in
creative ways.
The current work has in mind relatively sophisticated agents with
their own ``contextual maps'' and the ability to participate in
``reading and writing computational ecosystems''
\cite{antunes2015writing}.  Such agents will use, view, critique, and
evaluate the work and workflow of other agents.  Although
computational agents with all of these features do not exist yet in
any robust form, we can reason about them, and in so doing, help
design the future of \emph{computational social creativity}
\cite{saunders2015computational} -- an emerging research area at the
nexus of artificial life, social simulation, and computational
creativity.

\paragraph{Elinor Ostrom's ``design principles''}

To contextualise this effort, we must begin with a short excursus into
economics.  Ostrom's work is typically applied to study the management
of natural resources.  In economics jargon, the specific resources
considered are \emph{rivalrous} and \emph{non-excludable}.  This means
that consumption by one party precludes consumption by a rival, and
that it is not directly possible to for anyone to block others' access
to the resource.  Economic goods with these two properties are
referred to as \emph{common pool resources} (CPRs); see
\cite{ostrom2008challenge}.  Fisheries and forests are important
examples.  Economic actors have incentives to exploit these resources,
however, there are natural limits on total consumption.
In principle, a CPR might be gobbled up due to
individual greed: this is the so-called tragedy of the commons, and
one does not have to look too far for examples.  However, in practice,
the tragic outcome does not always transpire.  Ostrom's theoretical
and empirical perspective helps understand why, and emphasises:
\begin{quote}
(1) the importance of group attributes and institutional arrangements
  in relation to the structure of incentives and utilities for
  individual decision making; and (2) the likelihood of a broader set
  of possible outcomes, including user-group institutional solutions
  \cite[p.~23]{mccay1990question}
\end{quote}

Ostrom's ideas have recently been applied to analyse Wikipedia,
considered as an ``expressive commons''
\cite{safner-expressive-commons}.  Wikipedia is \emph{non-rivalrous}
in consumption, if we accept the metaphor ``to read is to consume.''
However, \emph{contribution} to Wikipedia presents a range of salient
social dilemmas, and efforts to manage them are reflected, for
example, in the \emph{Neutral Point Of View} (NPOV) policy, which
helps produce ``articles that document and explain major points of
view, giving due weight with respect to their prominence in an
impartial
tone.''\footnote{\url{https://en.wikipedia.org/wiki/Wikipedia:Five_pillars}}



IAD focuses on \emph{action situations}, framed in three phases:
\emph{context}, \emph{action}, and \emph{outcome}. 
Importantly, this part of the theory is not linked to
the particular details of CPRs.
Ostrom uses the term \emph{institution} to refer to the ``shared
concepts used by humans in repetitive situations organized by rules,
norms, and strategies'' \cite{ostrom2010institutional}.  We will
return to these concept categories later and consider them further
from a computational perspective.  For now, our way into thinking in
terms of IAD will be by way of several \emph{design principles} for
the successful management of CPRs that Ostrom described; see Table
\ref{tab:design-principles}.  These principles work together to
support institutions that maintain the integrity of the commons -- for
example, by ensuring that behaviour is monitored, that knowledgeable
and concerned parties are the ones who make specific rules, and that
conflicts do not get out of hand \cite[p.~79]{ostrom2012future}.





\paragraph{The four Ps}  We can bootstrap our contextual
understanding of creativity with the help of an existing model.
\citename[pp.~307-309]{rhodes1961analysis} intends ``the four Ps'' to
refer to the following facets of creativity, which are familiar from
everyday experiences of creativity in society.

\begin{itemize}
\item[] Person -- \emph{personality, intellect, temperament, physique, traits, habits, attitudes, self-concept, value-systems, defense mechanisms, and behavior.}
\item[] Process -- \emph{motivation, perception, thinking, and communication.}
\item[] Product -- \emph{an idea embodied into a tangible form.}
\item[] Press -- \emph{the relationship between human beings and their environment.}
\end{itemize}

We will shortly use these concepts to rewrite the items in Table
\ref{tab:design-principles}, replacing the focus on
\emph{appropriation} with a focus on \emph{contribution} that befits a
theory of social creativity.

\citeauthor{jordanous2016four} makes a case for thinking about
computational creativity using Rhodes's 4P's, starting with a critique
of the strategies used in the evaluation of computational creativity, which,
she emphasises, is ``traditionally considered \ldots\ from

\begin{threeparttable}[t]
\hspace{-.6em}\begin{minipage}{\columnwidth}
\bgroup
\def\arraystretch{1.2}%
\begin{tabular}{p{\columnwidth}}
\multicolumn{1}{c}{\textbf{Ostrom's design principles}} \\
\textbf{1A.}~\textbf{User boundaries}\newline\hspace*{.5cm}``Clear boundaries between legitimate users and nonusers must be clearly defined.''\tnote{2}\\
\textbf{1B.}~\textbf{Resource boundaries}\newline\hspace*{.5cm}``Clear boundaries are present that define a resource system and separate it from the larger biophysical environment.'' \\
\textbf{2A.}~\textbf{Congruence with local conditions}\newline\hspace*{.5cm}``Appropriation and provision rules are congruent with local social and environmental conditions.'' \\
\textbf{2B.}~\textbf{Appropriation and provision}\newline\hspace*{.5cm}``The benefits obtained by users from a common-pool resource (CPR), as determined by appropriation rules, are proportional to the amount of inputs required in the form of labor, material, or money, as determined by provision rules.'' \\
\textbf{3.}~\textbf{Collective-choice arrangements}\newline\hspace*{.5cm}``Most individuals affected by the operational rules can participate in modifying the operational rules.'' \\
\textbf{4A.}~\textbf{Monitoring users}\newline\hspace*{.5cm}``Monitors who are accountable to the users monitor the appropriation and provision levels of the users.'' \\
\textbf{4B.}~\textbf{Monitoring the resource}\newline\hspace*{.5cm}``Monitors who are accountable to the users monitor the condition of the resource.'' \\
\textbf{5.}~\textbf{Graduated sanctions}\newline\hspace*{.5cm}``Appropriators who violate operational rules are likely to be assessed graduated sanctions (depending on the seriousness and context of the offense) by other appropriators, by officials accountable to these appropriators, or both.'' \\
\textbf{6.}~\textbf{Conflict-resolution mechanisms}\newline\hspace*{.5cm}``Appropriators and their officials have rapid access to low-cost local arenas to resolve conflicts among appropriators or between appropriators and officials.'' \\
\textbf{7.}~\textbf{Minimal recognition of rights to organise}\newline\hspace*{.5cm}``The rights of appropriators to devise their own institutions are not challenged by external governmental authorities.'' \\
\textbf{8.}~\textbf{Nested enterprises}\newline\hspace*{.5cm}``Appropriation, provision, monitoring, enforcement, conflict resolution, and governance activities are organised in multiple layers of nested enterprises.'' 
\end{tabular}
\egroup
\end{minipage}
\vspace{-.3cm}
\caption{Ostrom's design principles as expressed in the meta-review carried out by \citename{cox2010review} \label{tab:design-principles}}
\end{threeparttable}

\vspace{.1cm} \setcounter{footnote}{2} \footnotetext{Repetition is
  \emph{sic}, the point being that the boundaries must be both
  distinct and explicitly defined.}


\noindent the perspective of the creative output produced by a system''
\cite{jordanous2016four}.

Ecological thinking suggests that that it is quite limited to take the
final product as the sole term of analysis.  At least we might like to
introduce the ``embedded evaluation'' of creative products into the
creative process, and build agents that are aware of some contextual
features of their environment.  For example, these agents might ask:
\emph{How similar or how} \emph{different is my generated artwork to
  an existing artwork, or to the components thereof, or to the initial
  conception for the work?}  This route is quite close to Ritchie's
\shortcite{ritchie07} empirical criteria for judging a final product
against an ``inspiring set'' -- but now makes evaluation an explicit
part of the creative process.  Some recent work in computational
creativity emphasises embedded evaluation \cite{gervas2014reading}.
However, as Jordanous argues, creative products are just one part of
the overall creative process -- and the 4Ps help expose the other features.

\begin{table}[t]
\bgroup
\def\arraystretch{1.2}%
\hspace{-.6em}\begin{tabular}{p{\columnwidth}}
\multicolumn{1}{c}{\textbf{Proposed creativity design principles}}\\
\textbf{1A.}~The population of \Person[s]\ who can add to or alter the resource is clearly defined.\\
\textbf{1B.}~The boundaries of the \Place\ must be well defined.\\
\textbf{2A.}~The \Process\ is related to local conditions.\\
\textbf{2B.}~Contributing to the \Product\ has benefits for the \Producer\ that are proportional to the efforts expended.\\
\textbf{3.}~Most \Producer[s]\ who are affected by the rules governing contribution can participate in modifying the operational rules. \\
\textbf{4A.}~Tests document the interaction of \Producer[s] and \Place. \\
\textbf{4B.}~Tests can be modified by \Producer[s] or their representatives. \\
\textbf{5.}~\Producer[s]\ who violate operational rules in the domain will be assessed sanctions by other \Producer[s].\\
\textbf{6.}~\Producer[s]\ have rapid access to low-cost local arenas to resolve conflicts. \\
\textbf{7.}~The rights of \Producer[s]\ to devise institutions governing their contributions are not challenged by external authorities.\\
\textbf{8.}~Contribution, testing, enforcement, conflict resolution, and governance and are organised in multiple layers of nested \Place[s]\ and agencies.
\end{tabular}
\egroup
\caption{``Creativity design principles'' formed by switching the polarity of entries in Table \ref{tab:design-principles} to emphasise contribution rather than appropriation, and using the concept of ``tests'' to connect to computing practice \label{tab:creativity-design-principles}}
\end{table}


Unfortunately, however, Rhodes's thinking and terminology is too
anthropocentric for our current purpose.  As Ostrom describes it,
action situations are to be understood using seven clusters of
variables: \emph{participants}, \emph{positions}, \emph{potential
  outcomes}, \emph{action-outcome linkages}, \emph{participant
  control}, \emph{types of information generated}, and \emph{costs and
  benefits assigned to actions and outcomes}
\cite[p.~14]{ostrom2009understanding}.  Nowhere does this mention a
``Person''.  Continuing the adaptations begun by
\citename{jordanous2016four}, the four Ps will be rendered here as
\Producer\slash\Process\slash\Product\slash\Place.  It is important to
emphasise that these labels are strictly more inclusive than Rhodes's,
and more abstract.  In particular, the \Place\ corresponds to Ostrom's
action situation, structured in advance by contextual features.
This adapted 4P model is reminiscent of the
\emph{Domain-Individual-Field Interaction} (DIFI) model due to
\citename{feldman1994changing}, if we understand Domain
$\approx$ \Place, Individual $\approx$ \Producer, and Field $\approx$
(a collection of) established \Process[es].
Note that contextual theories, broadly construed, pose a long-standing
challenge for computing, partly because ``what context is changes with
its context'' \cite[p.~343]{gundersen2014role}.  One possible working
definition is that: ``Context is what contrains a problem solving
[scenario] without intervening in it explicitly''
\cite{brezillon1999context}.  Another relevant remark is that context
is ``defined solely in terms of effects in a given situation''
\cite{hirst1997context}.

In developing an \emph{institutional approach to computational social
  creativity}, we will look for the rules, norms, and strategies that
can be used to establish suitable and effective contextual
relationships between \Process[(es)], \Place[(s)], \Producer[(s)], and
\Product[(s)].

\paragraph{Transposing the design principles into ``creativity design principles'' and translating them into technical terms}

\emph{Software testing} is embodied in the formal ideas of
\emph{assertions}, \emph{advice}, and \emph{contracts}.  Related
programming methodologies aim to build \emph{executable
  specifications} and may make use of \emph{test-driven development}
(TDD).  These techniques provide various ways for (evolving) programs
to interact with their context.  These ideas can help us translate
Table \ref{tab:design-principles} into technical terms.
To get started, Table \ref{tab:creativity-design-principles} uses the
4P terminology and the generic notion of a test to transpose Ostrom's
design principles into ``creativity design principles.''

\section{Testing for creativity}

The current section elaborates the candidate creativity design
principles outlined above, expanding each with relevant literature and
examples, and seeking the ways in which each principle could be
applied within a software system.





\paragraph{1A.~User boundaries} ~

``The population of \Producer[s]\ who can add to or alter the resource is clearly defined.''

In user-oriented computing, this principle is often addressed using
\emph{Access Control Lists} (ACLs) or other permissions mechanisms.
The corresponding tests are relatively simple: either each modifiable
object in the system has a piece of metadata about it that says
who can modify it, or each user has a piece of metadata attached to
his or her user account that says which resources they can modify.

Before granting access to a resource, we may require that a
\Producer\ implements certain protocols.  
In a client-server architecture, the client
generally communicates using an existing API and may need to
implement a certain set of callback functions or adhere to other
restrictions.
Noncompliant user behaviour after access has been granted may result
in access being revoked.  Thus, for example, even though Wikipedia is
``the encyclopedia anyone can edit,'' violating the site's principles
may lead to a IP-based block, or a username-based ban.

\paragraph{1B.~Resource boundaries} ~

``The boundaries of the \Place\ must be well defined.''

The source of this well-definedness may come from ``either
side.''  That is, the \Place\ may advertise its definition in terms of
its APIs and other criteria (as above) together with guarantees
on output behaviour in the style of ``Design by Contract''
\cite{mitchell2002example}; alternatively, \Producer[s]\ may implement
tests that restrict the \Place[s]\ that they will engage with.

In a simple example of the latter sort, a game-playing agent might
resign if it estimates that its position is unwinnable.
The fact that different participants can have different perspectives
points to an interesting special case in which the (shared) definition
of the \Place\ arises in an emergent manner.  This phenomenon
is especially important if we ``[take] a broad view of creativity as any process
in which novel outcomes emerge'' \cite{saunders2015computational}.

\paragraph{2A.~Congruence with local conditions} ~


``The \Process\ is related to local conditions.''

In Ostrom's original formulation, local conditions were broken down
along axes of ``time, place, technology, and/or quantity of resource
units'' \cite{elinor1990governing}.
With respect to theorising the local conditions of creativity, we can
gain a useful perspective by turning briefly to the psychoanalyst
Winnicott's treatment of ``the exciting interweave of subjectivity and
objective observation'' which takes place in an ``area that is
intermediate between the inner reality of the individual and the
shared reality of the world'' \cite[p.~86]{winnicott1971playing}.
We are then led to consider those local conditions that exist in the
``interweave'' of \Place\ and \Producer.  For example, roboticist
Andy Clark proposes a theory of extended cognition, in which enminded
beings ``use'' the environment to self-program and are not
just programmed by the environment \cite{clark1998being}.  However as Clark
points out elsewhere,
``it becomes harder and harder to say where the world stops
and the person begins'' \cite{clark2001natural}.
%
In short, the mind is not separated from the body or environment but
grounded in perception \cite{ingold2000perception}.

A corresponding computational test is found in the earlier example of
embedded evaluation, in which existing artefacts are employed as a
virtual sensorium.  More broadly, this principle concerns making sense
of, or ``parsing'', the \Place\ (and the other P's).  This 
\Process\ is well described by Steigler's notion of
\emph{grammatisation}: ``processes by which a material, sensory, or
symbolic flux becomes a gramme,'' or, more simply, ``the production
and discretisation of structures'' \cite{tinnell2015grammatization}.
Remember that while a given agent is trying to make sense of the
world, others are likely trying to make sense of that agent as well.  Framed as
dilemma, the last word would likely be: ``program or be programmed''
\cite{rushkoff2010program} -- but reflecting on Clark's comment above,
we see that this can become somewhat complex.

\paragraph{2B.~Appropriation and provision} ~

``Contributing to the \Product\ has benefits for the \Producer\ that are proportional to the efforts expended.''

The usual way of thinking about computers -- as non-agentive machines
-- would render the above-stated principle perfectly
meaningless.  In connection with principle 2A, we should here remark:
``That an object is more profitable or effective is only a secondary
consequence of its refinement'' \cite[p.~12]{chabot2013philosophy}.
In any case, before we can think about ``benefits'' in the case of a
non-human (and non-living) \Producer, the phrasing of the
current principle leads us to ponder the cost of their ``efforts.''

It may be best to change tack, and ask, with Terrance Deacon, ``In
what sense could a machine be alive?''~\cite{machine-alive}.  If a
machine were responsible for maintaining its own energy supply, its
features of outward-orientation might give cause to say that the
machine has a ``self'' \cite{deacon2011emergence}.  Consider for
example the Ethereum project, which provides protocols for distributed
computing and the creation of ``decentralized autonomous
organisations'' -- whose organisation relative to the outside world is
mediated by cryptocurrency, referred to as ``fuel''
\cite{wood2014ethereum}.


From a testing standpoint, the key requirements are: an ability to
judge whether a given option can be (tentatively) thought of as
beneficial, and, ideally, a memory that can compare these judgements
with iterations of similar situations later on.  In this way we
would recover the foundations of reinforcement learning, and, as
Ostrom points out, the core logic behind the development of new
institutions:
\begin{quote}
``How about if you do $A$ in the future, and I will do $B$, and before we ever make a decision about $C$ again, we both discuss it and make a joint decision?'' \cite[p.~19]{ostrom2009understanding}
\end{quote}

\paragraph{3.~Collective-choice arrangements} ~

``Most \Producer[s]\ who are affected by the rules governing contribution can participate in modifying the operational rules.''

Let us reflect in more detail on the \emph{rules} that comprise -- along with
biophysical and material conditions and community attributes -- 
the locally-contextual variables which determine or constrain an
action situation \cite[p.~15]{ostrom2009understanding}.  At their
simplest, these rules are ``if-then'' statements giving instructions
that determine the behaviour of persons in certain roles.  As such,
each rule contains a logical test, and changing the rules
means writing new tests.

Ostrom develops a grammar around this idea, and defines \emph{regulatory rules} with
the following formula:

\begin{quote}
ATTRIBUTES of participants who are OBLIGED, FORBIDDEN, OR PERMITTED to
ACT (or AFFECT an outcome) under specified CONDITIONS, OR ELSE. \cite[p.~187]{ostrom2009understanding}
\end{quote}

\emph{Norms} and \emph{strategies} are defined using a simplified
formula, also cast in terms of \emph{attributes},
\emph{deontics},\footnote{I.e., the presciptive valence -- obliged,
  forbidden, or permitted, as above -- for norms, not for
  strategies.} \emph{aim}, and \emph{conditions}
\cite[p.~140]{ostrom2009understanding}.  The prescriptive terms may be
assigned a particular weight, and actions and consequences may also be
assigned a particular cost or value
\cite[p.~142]{ostrom2009understanding}.  Some relevant actions are:
\emph{be} in a position, \emph{cross} a boundary, \emph{effect} a
choice, \emph{jointly exercise} partial control together with others,
\emph{send or receive} information, \emph{pay out or receive} costs or
benefits, and \emph{take place} (for outcomes)
\cite[p.~191]{ostrom2009understanding}.

Something more needs to be said about the assertion that
\Producer[s]\ ``can'' participate in changing (or creating) rules,
norms, and strategies.  In practice, participatory systems tend to be
lossy.  Changes to rules and structures will tend to be carried out by those
\Producer[s]\ who are \emph{most} affected -- and \emph{thus} most knowledgeable;
cf.~\citename[p.~79]{ostrom2012future}.  The structure
of new rules is predicted by Conway's Law:
\begin{quote}
[T]here is a very close relationship between the structure of a system and the structure of the organization which designed it.  \cite{conway1968committees}
\end{quote}

Specifically, the proposed relationship is ``homomorphism'':
following Conway, any \Product\ will mirror the hyper-local conditions
that describe the \Producer[s]' social context.  Furthermore, it seems
likely that \Product[s]\ will mirror local environmental conditions in
the \Place.
This points to importance of a broad class of tests that would be
described as environmental ``sensors''.  This theme will be developed
more fully below.


\paragraph{4A.~Monitoring users} ~

``Tests document the interaction of \Producer[s] and \Place.''

The straightforward view suggested by the idea of ``monitoring'' is to deploy some global functionality that keeps track of the actions of all participating \Producer[s] within a \Place.  But this function can be broken up and distributed out among the \Producer[s] themselves. In the first instance, what a \Producer\ produces is sensory data. Sensors are generally deployed along with effectors or (more broadly) transducers that translate the sensory information into action. So, monitoring is important for modelling any action or interaction whatsover.  For example, The Painting Fool compares an initial altered snapshot (sensory data) to the painted image that it generates in response to that snapshot, and judges the quality of its output on that basis \cite{colton2014you}.  This example could be extended to theorise ``proprioceptive'' sensing and judgement about effected actions more broadly. Filtering upstream data is another simple application of sensors, which \citename{keller2012sonic} describes as an ``ecocompositional technique.''  In short, an ecological view on monitoring suggests that it can be distributed out among participants and that this is vital for social creativity.


\paragraph{4B.~Monitoring the resource} ~

``Tests can be modified by \Producer[s] or their representatives.''

The environment itself also filters and selects \cite{kockelman2011biosemiosis}.  Some of these conditions are fatal for living beings in the environment, and more broadly may provide terminating conditions for the constituent \Process[es] in a \Place.  It would be too much to say that \emph{all} tests can be modified by \Producer[s].  Rather, \Producer[s] may have programmatic access to those tests which transform potentially fatal (or at least fateful) features of the \Place\ and participating \Producer[s] into \emph{data}.  This opens up the possibility of directly modifying decision making processes on the one hand, or of passing along information about the fitness landscape to future generations of \Producer[s] in a (co-) evolutionary framework on the other
\cite{delanda2011philosophy}.

Simply put, \emph{data} is lack of uniformity within some context \cite{sep-information-semantic}.  In the case of monitoring the extractive use of CPRs, direct and compelling feedback about instances of non-uniform or otherwise aberrant resource usage define critical (i.e., decisive) points within a resource management structure.  In creative contexts ``critique'' is no less important.

\paragraph{5.~Graduated sanctions} ~

``\Producer[s]\ who violate operational rules in the domain will be assessed sanctions by other \Producer[s].''

Economic sanctions are generally punishments, which are presumed to have a clear meaning or a direct impact on behaviour.  However, there are other cases in which \Producer[s]' interactions with other \Producer[s] will not be punitive so much as, for example, educative or otherwise formative.  

In an artistic context, ``sanctions'' may range from constructively
critical reviews to outright condemnation to no response at all.
The \emph{Iterative Development-Execution-Appreciation} (IDEA) cycle
\cite{colton2011computational} introduces \emph{well-being} and
\emph{cognitive effort} ratings from which several derived measures of
audience response can be computed (e.g., by averaging across audience
members).  This can readily be extended to a developmental or peer
production context.  ``Audience'' might be re-thought as a ``public,''
or as Rhodes's ``press'' (as originally formulated) to
capture the idea that its response has a direct effect on the
\Producer.  Inasmuch as the \Producer\ is produced, feedback from the
``parent'' \Producer[(s)]\ is especially important to this formative
\Process.


\paragraph{6.~Conflict-resolution mechanisms} ~

``\Producer[s]\ have rapid access to low-cost local arenas to resolve conflicts.''

Wikipedia's edit wars provide a familiar example
\cite{viegas2007talk,yasseri2012dynamics}.  These are carried out on
the pages of the encyclopedia itself, and resolved using supplementary pages.
Machine-generated metadata is relied upon throughout.  These
mechanisms are low cost: the ``stigmergic'' self-organisation patterns
exemplified by open online communities make fairly minimal demands on
participating agents \cite{heylighen2015stigmergy}.  Nevertheless,
structure matters: cases of direct and unresolvable conflict must
usually be referred a higher authority, e.g., sitewide
guidelines and policies, or available arbitration committees. 
Opportunities to jointly exercise partial control are, again,
often \Product[s], and the creation of a communication channel
-- a \Place\ within a \Place\ -- is another formative \Process,
which \citename[p.~355]{jakobson1960linguistics} calls the ``phatic function.''

The theme of local scale suggests more and less representative
examples.  For instance, academic research is currently organised in a
much more segmented and localised format than Wikipedia.  
\emph{Modularity} is one of three features that are hypothesised to
support \emph{commons based peer production} (CBPP)
\cite{coases-penguin}.  However, CBPP requires not just
decomposability into modules but relatively fine \emph{granularity} of
these modules, and as well as a \emph{low cost of integration} to
bring disparate pieces of work together once they are completed --
possibly ``subsidised'' by an assistive technology, like Wikipedia's
metadata systems.  Creative and scientific writing, at the level of
individual papers or books, tends to miss features that would allow
this work to scale up \cite{kim2014ensemble} -- even though science
and literature represent impressively huge ``virtual'' collaborations.

The most straightforward test related to this theme is that a
\Producer\ needs to be able to \emph{detect} conflict, either between
itself and other \Producer[s], or between incompatible goals.
In order to resolve a conflict -- or to organise work on a project to
avoid conflicts in the first place -- a \Producer\ will probably need
to reason about the project's structure.

\paragraph{7.~Minimal recognition of rights to organise} ~

``The rights of \Producer[s]\ to devise institutions governing their contributions are not challenged by external authorities.''

The foremost external authority to be concerned about in a
computational creativity setting is the programmer.  A ``mini me''
critique can readily be levelled by CC sceptics
\cite{colton2012painting}.  We are still in early days for autonomous
creative systems and general AI, and involvement of programmers and
others in teaching systems how to devise institutions is
at least as relevant as teaching them how to conform to pre-given instructions.

Keeping in mind the earlier reflections on Winnicott, a relevant set
of tests would compare the frequency of user- or programmer-generated
changes in the system, with the frequency of changes coming from the
system itself.  This is the thrust of the diagrammatic formalism of
creative acts developed by \citename{colton-assessingprogress}: with
considerable further work we could expand the ability of computer
systems to participate in, or fully automate, such modelling activity.
A basic challenge in applying the formalism from
\citeauthor{colton-assessingprogress}~is to identify the individual
``creative acts'' that a given \Producer\ has made.  The tests that
would reveal these acts in a given stream of \Product[s]\ tend to be
domain-specific.

\paragraph{8.~Nested enterprises} ~

``Contribution, testing, enforcement, conflict resolution, and governance and are organised in multiple layers of nested \Place[s]\ and agencies.''

That the \Place\ or the \Producer\ would be layered isn't a 
surprise; many systems have a hierarchical aspect.  
What is perhaps more surprising is that many of the features
that make up a ``creative ecosystem'' must themselves
be \emph{produced}, which points to the inherent
multiplicity of \Producer[s]. 
Here, \Producer[s]\ are seen as self-organising the structure of
their interrelationships and interconnections at various levels.
Developing a computational treatment of such a system divorced from
real world applications would be a thankless and ultimately futile
task.  Effort may be better spent on developing programs that model
and participate in existing creative ecosystems.
In such cases, there would be real-world empirical tests of success,
coming from users.

\section{Example}

This section uses the creativity design principles discussed above to
describe some of the creativity-supporting institutions in place at
the Seventh International Conference on Computational Creativity (ICCC 2016),
and to explore potential additions and adaptations for future ICCCs.

\medskip

\noindent \textbf{I.}  The crucible for the current paper was a unique
set of ongoing discussions (see ``Acknowledgements'') (2A).
At first, the hope was to co-author the paper with one of these
discussants, but due to time constraints this was not possible, so it
became a single-author paper (1A).  The ICCC call helped motivate
writing up the ideas (2B), partly because the conference is open to
papers that are informed by and contribute to various disciplines
at varying degrees of formality.  However, ICCC enforces rigourous
academic standards, using slightly different evaluation criteria for
papers submitted to each of five ``tracks'' (1B).
Reviewers used the Easychair website to bid for papers to review, and
to share discussions and debate about these papers in case of
disagreement (6).  Papers that were seen as less relevant were
rejected outright, or potentially (as a norm) allocated briefer slots
in the conference schedule (5).  The current paper was conditionally
accepted, which meant that it entered into a ``shepherding'' process,
whereby a senior programme committee member could check (4A) whether
the author followed through on specific reviewer requirements (4B).
By and large authors are given free rein to write papers about any topic
relevant to computational creativity, if they do so in a rigorous
academic style (7).  This entails reflecting on certain
themes-held-in-common -- but the conference seems to lose some
opportunities for structuring engagement more deeply, e.g., around
common tools or challenge problems (8?).  Presumably only the
conference steering committee can change the conference's overall
rules; however, it should be noted that reviewer requirements
constitute fine-tuned rule-setting at the level of individual papers
(3?).

\medskip

\noindent \textbf{II.}  The reflections above begin to suggest ways in
which we might make better use of software systems in creative
partnership. 
One realistic idea
would be to use computer programs to help with paper review tasks.
Essay grading software is now mainstream, and services like WriteLab
can help authors simplify their writing and catch grammar and logic
errors.\footnote{\url{https://writelab.com} offers a freemium service
  for students, but is ``always free for instructors.''}  Agent-based
reviews or a shift to \emph{post-publication review}, in which reviews
are offered ``after an article is published, much like commentary on a
blog post'' \cite[p.~316]{ford2013defining} would change the
population of reviewers (1A).  Moving beyond blogs to wikis, lists of
open problems from prior publications could be collected, compared,
and explicitly referenced with semantic links (1B)
\cite{tomlinson2012massively}.  This could begin to make explicit the
ways in which a given paper constitutes an advance (2A, 2B).  The
development, use, and maintenance of shared tools (APIs,
open source software) and design patterns for computational creativity
could be encouraged (3).  A standardised testing approach based on
challenge problems, as in the recently announced OpenAI
Gym,\footnote{\url{https://gym.openai.com/}} with worked examples,
explicit evaluation metrics, and variant versions (4A, 4B) could help
the community move towards, and enforce, standards of
\emph{replicability} and \emph{generalisability} (5).  Partial
``wikification'' and semantisation of the research area is already
underway with systems like FloWr \cite{charnley:iccc2014} and
ConCreTeFlows \cite{concreteflowsreport},
but it is unclear whether these systems will merge, or diverge, or if
a new standard will come along (6?).  Once shared technologies and
datasets are in common use, computational agents will be better able
to contribute to the field (7).  It is to the advantage of
computational creativity researchers to develop applications and
application environments that we -- and others -- agree are useful
(8).

\section{Discussion and Conclusions}

This section reviews the contribution above, beginning with a link to
related work.  Specifically, Ostrom's high-level Institutional
Analysis and Development (IAD) framework can be regarded in parallel
with the high-level outline of the \emph{Standardised Procedure for
  Evaluation of Creative Systems} (SPECS) \cite{jordanous:12}.  SPECS
suggests that, in order to evaluate creativity, it is necessary to put
forth a definition of ``what it is to be creative,'' and then to
specify criteria by which creativity will be measured before
formulating the evaluation.  IAD suggests that institutions operate
within a certain context, which afford certain kinds of actions, and
that these lead to certain observable outcomes. To wit:

\vspace{-1mm}
\begin{center}
\begin{tabular}{cc}
\textbf{IAD} & \textbf{SPECS}\\
\begin{tabular}{|l|}
\hline
Context\\ \hline
Action\\ \hline
Outcome\\ \hline
\end{tabular}&
\begin{tabular}{|l|}
\hline
Definition\\ \hline
Criteria\\ \hline
Evaluation\\ \hline
\end{tabular}
\end{tabular}
\end{center}

In IAD, context can be thought of as a collection of ``exogenous
variables,'' \cite[p.~13, esp.~Figure~1.1]{ostrom2009understanding} including pre-defined rules, that shape what happens in
the action situation at the heart of the analysis.  We have described
several candidate design principles that outline
potential rules for guiding action in creative settings.  
This suggests the possibility of recording
a definition and set of criteria for evaluating social creativity
in a general domain.  Pragmatically, this definition might
unpack the 4Ps in terms of Ostrom's variables (participants, positions, etc.).

SPECS could be criticised for being overly abstract: in other words,
for simply describing good practice in any empirical investigation.
IAD adds many more specifics, which have necessarily been presented in
a compressed form here.  It is hoped that this first attempt to use IAD
to theorise computational social creativity will motivate future
explorations that further unpack social creativity using Ostrom's ideas.


The creativity design principles offer guidelines (and with minor
changes, hypotheses) for members of the computational creativity community
to test out in practice.
More empirical work is needed to validate (or improve) these
principles.  On the cultural side, more attention should be given to
the fact that our institutions -- including institutions for building institutions
-- are analysable in programmatic terms.
%


\section{Acknowledgements}
This work was supported by the Future and
Emerging Technologies (FET) programme within the Seventh Framework
Programme for Research of the European Commission, under FET-Open
Grant number 611553 (COINVENT).

Ongoing conversations with Anna Jordanous, Steve Corneli, and Rasmus
Rebane about computational creativity, economics, and semiotics
(respectively) were key drivers in this paper's evolution.  The
anonymous reviewers made many useful comments on matters of presentation.

\bibliographystyle{iccc}
\bibliography{iccc}

\end{document}